\documentclass[preprint,pra,nofootinbib]{revtex4}
\usepackage[mathscr]{euscript}
\usepackage{graphicx}
\usepackage{float}
\usepackage{dcolumn}
\usepackage{bm}
\usepackage{graphicx}
\usepackage{amsmath,amssymb,amsthm}
\usepackage{subfigure}
\usepackage{color}
\usepackage{hyperref}
\hypersetup{colorlinks=true,linkcolor=blue}

\newcommand{\bea}{\begin{eqnarray}}
\newcommand{\eea}{\end{eqnarray}}
\newcommand{\beq}{\begin{equation}}
\newcommand{\eeq}{\end{equation}}

\def\/{\over}

\newcommand{\rev}[1]{\textcolor{black}{{#1}}}

\begin{document}

\title{Observation of localization of light in photonic quasicrystals \rev{of diverse symmetries}}

\author{ Peng Wang,${}^{1}$ Qidong Fu,${}^{1}$ Vladimir V. Konotop,${}^{2}$ Yaroslav V. Kartashov,${}^{3}$ Fangwei Ye$^{1\ast}$\\
\emph{${}^{1}$School of Physics and Astronomy, Shanghai Jiao Tong University, Shanghai 200240, China}\\
\emph{${}^{2}$Centro de F\'{i}sica Te\'{o}rica e Computacional and Departamento de F\'{i}sica, Faculdade de Ci\^encias, Universidade de Lisboa, Campo Grande, Ed. C8, Lisboa 1749-016, Portugal}\\
\emph{${}^{3}$Institute of Spectroscopy, Russian Academy of Sciences, Troitsk, Moscow, 108840, Russia}\\
\emph{$^\ast$Corresponding author: fangweiye@sjtu.edu.cn}
}

\date{\today}


\maketitle

\textbf{Since their first observation in metallic alloys \cite{Shechtman1984} and until now, quasicrystals \cite{Levine1984} remain among the most intriguing physical structures sharing properties of ordered and disordered media. Quasicrystals are ubiquitous in nature. Beyond crystalline solids~\cite{Steurer2018,Steurer2004}, they can be created as optically induced or technologically fabricated structures in photonic \cite{Freedman2006,Vardeny2013} and phononic \cite{Steurer2007} systems, as potentials for cold atoms \cite{Guidoni1997} and Bose-Einstein condensates (BECs)~\cite{Sanchez2005}. On a par with the unusual structural properties of quasicrystals, nowadays the problem of wave propagation in such two-dimensional structures attracts considerable attention, with many aspects requiring exploration due to strikingly different localization properties observed in various quasicrystalline systems. Already in earlier studies \cite{Kohomoto1986} it was predicted that the lowest electronic states in \rev{five-fold} quasicrystals are localized. \rev{Later on localization of BECs in an eight-fold rotational symmetric quasicrystal optical lattices was observed~\cite{Sbroscia2020}. Meantime}, experimental studies of photonic quasicrystals demonstrate localization of light and suppression of transport only in the presence of nonlinearity \cite{Freedman2006} or additional disorder \cite{Levi2011}. Direct observation of localization in purely linear \rev{photonic} quasicrystals, therefore, remains elusive \rev{, and further, the impact of the varying rotational symmetry on the localization is yet to be understood.} Here, using sets of interfering plane waves, we create photonic two-dimensional quasicrystals with different rotational symmetries, not allowed in periodic crystallographic structures. We demonstrate experimentally that linear localization of light does occur even in clean linear quasicrystals for probe beams propagating both in the center and off-center regions of the quasicrystals. We found that light localization occurs above a critical depth of optically induced potential and that this critical depth rapidly decreases with the increase of the order of the discrete rotational symmetry of the quasicrystal. Our results clarify a long-standing problem of wave localization in linear quasicrystals and elucidate the conditions under which this phenomenon occurs. The localization, that we observe here for different symmetries of quasicrystals and for excitations in different spatial locations, is thus shown to be a robust phenomenon. These findings pave the way for achieving wave localization in a wide variety of aperiodic systems obeying discrete symmetries, with possible applications in photonics, atomic physics, acoustics, and condensed matter.}

Quasicrystals~\cite{Shechtman1984,Levine1984,Steurer2018,Steurer2004} are unique structures: unlike crystals, they are not periodic, i.e., they do not feature translational symmetry, and at the same time, they still can continuously fill the entire space. Unlike crystals, which by the crystallographic restriction theorem can possess only two-, three-, four-, and six-fold rotational symmetries, two-dimensional quasicrystals can feature any order of a discrete rotational symmetry, like for example, five-fold, seven-fold, or higher symmetries (see examples in Fig.~\ref{fig:one} below).
Quasicrystalline structures, initially discovered in the process of the growth of alloys~\cite{Steurer2004, Steurer2018}, are nowadays extensively studied in solid-state physics ~\cite{Shechtman1984, Steurer2004, Steurer2018, Kamiya2018} including twisted bilayered graphene~\cite{Ahn2018, Yao2018}. They are artificially created in ultracold quantum gases ~\cite{Sanchez2010, Mace2016, Lellouch2014, Sbroscia2020}, in various optoelectronic ~\cite{Tanese2014,Goblot2020} and photonic ~\cite{Chan1998,Freedman2006,Freedman2007,Xavier2010,Boguslawski2011} systems. Among the most interesting aspects of quasicrystalline materials that are under active current investigation in diverse fields are the impact of symmetry and the long-range order of such structures on wave propagation in them.


Indeed, the evolution, transport, and localization properties of waves in a given medium are determined, in particular, by its geometrical characteristics and, specifically, by its inner symmetry. For example, localization of linear excitations is impossible in homogeneous and periodic media due to their translational symmetry. At the same time, linear localization is possible in two-dimensional disordered materials~\cite{disorder,disorder1}. More recently it was predicted~\cite{LocDeloc} and observed experimentally~\cite{Wang2020} that localization of light can also occur in incommensurate moir\'e lattices, created by two twisted periodic sublattices and characterized by rotational point symmetry, but lacking translational symmetry. However, in all previous experimental studies of light propagation in two-dimensional quasicrystals, the localization was reported only under the action of nonlinearity~\cite{Freedman2006} or of additional disorder ~\cite{Levi2011}. Thus, the observation of wave localization in pure quasicrystals without the action of additional confining factors remains elusive. The present work provides clear and direct experimental proof of this intriguing phenomenon by studying light propagation in reconfigurable photonic quasicrystals with various discrete symmetries. For the first time to our knowledge, the dependence of the depth of optical potential, at which light localization occurs, on the order of its rotational point symmetry is revealed. 

\begin{figure}[htb]
\centering
\includegraphics[width=\linewidth]{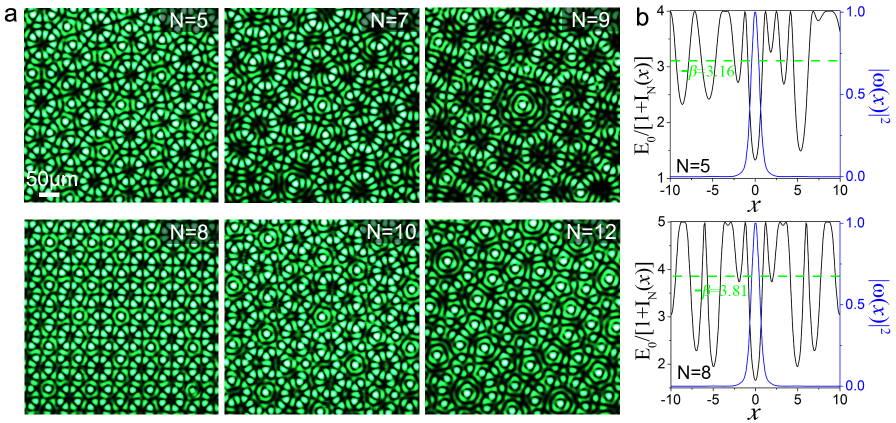}
\caption{\setlength{\baselineskip}{10pt} 
 {\bf Quasicrystals of different symmetries and cross-sections of corresponding optical potentials.} ({\bf a}) Experimentally generated $N$-fold symmetric quasicrystal patterns $I_N(\bf r)$ with odd (upper row) and even (lower row) dihedral symmetries. The reference origin is the image center.
 ({\bf b}) The representative numerically generated cross-sections of the optical quasicrystal potential $E_0/[1+I_N(\bf r)]$ for $N=5,E_0=4$ (upper panel) and $N=8,E_0=5$ (lower panel) along $x$ axis at $y=0$ of the lattice profile. The horizontal green lines in the right column show the "energies" equal to $-\beta$ for the fundamental, most localized modes shown in blue curves.
} 
\label{fig:one}
\end{figure}

We optically induce \cite{Freedman2006} photonic quasicrystals in a photorefractive crystal, SBN: 61 (SBN, strontium barium niobate) with dimensions 5 $\times$ 5 $\times$ 20 mm$^3$, by interfering $N$ pairs of counterpropagating coherent plane waves with transverse wavevector components $\pm \bm {k}_1, \pm \bm{k}_2, ..., \pm \bm{k}_N$  having the same absolute values and mutually rotated by the angles rotated $2\pi/N$, i.e.,  ${\bm k}_j=  k (\cos\theta_j, \sin\theta_j)$ with a constant $k$ and $\theta_j=2 \pi(j-1)/N$. SBN crystals feature a strong electro-optic anisotropy, with the electro-optic coefficient $r_{\text{13}}=~45~\text{pm}/\text{V}$  being substantially smaller than $r_{\text{33}}=~250~ \text{pm}/\text{V}$. Accordingly, we use ordinary polarized light (affected by $r_{13}$) for the lattice induction, so that the corresponding beams do not experience any noticeable self-action in the crystal and propagate undistorted as in uniform linear medium. In contrast, the probe light is extraordinary polarized (affected by $r_{33}$) so it feels inhomogeneous refractive index landscape induced by the ordinary polarized beam. 

In the paraxial approximation, the propagation of an extraordinarily polarized probe beam with dimensionless amplitude $\psi ({\bm r},z)$, in a photorefractive medium with an optically induced refractive index landscape is governed by the Schr\"{o}dinger equation~\cite{Freedman2006}:
\begin{equation}
	\label{equation1}
	\textit{i}\frac{\partial {\psi}}{\partial z}=H\psi , \qquad H=-\frac{1}{2}\nabla^2  +\frac{E_0}{1+I_{{N}}({\bm r})}.   
\end{equation}
Here $\nabla=(\partial/\partial x,\partial/\partial y) $; ${\bm r}=(x,y)$ is the radius-vector in the transverse plane scaled to the wavelength $\lambda=632.8~\textrm{nm}$ of the beam used in the experiments; $z$ is the propagation distance scaled to the diffraction length $2\pi n_{\rm e} \lambda$;  $n_{\rm e}$ is the refractive index of the homogeneous crystal for extraordinary-polarized light; $E_0>0$ is the dimensionless potential amplitude, controlled by bias field $E$ through   $E_0=k_0^2{n_e^4}D^2r_{33}E/2$. Here $k_0=2\pi/ \lambda$ is the wave number, and $D$ is the unit of the transverse distance.
The intensity of the $N$-fold symmetric optical lattice, which is induced in the sample by $N$ pairs of counterpropagating plane waves, is given by $I_{N}({\bm r}) = [
\displaystyle{(A/N)}
\sum_{j=1}^N \cos({\bm k}_j \cdot {\bm r}+\varphi)]^2 $.  The nonzero stationary phase $\varphi$ breaks the inversion ${\bm r}\to -{\bm r}$ symmetry, thus enabling rotational symmetries of odd orders. The amplitude of each plane wave is chosen such that the maximum of $I_{N}({\bm r})$ and consequently the maximal depth of the lattice potential, given by $E_0/[1+I_N(\bf r)]$, remains the same for all quasicrystals used in the experiment regardless of their rotational symmetry. In our simulations, $k=2$, $\varphi=\pi/10$, and $A^2=2.24$. Such photonic quasicrystals obey two-dimensional dihedral symmetries $D_N$, accounting for $N$ rotations and $N$ reflections. Respectively, the group properties differ for even and odd $N$. Restricting the consideration to non-crystalline symmetries, in Fig.~\ref{fig:one} we show examples of the experimentally created quasicrystals with $N=$5,\,7,\,9 (upper row) and $N=8,\,10,\,12$ (lower row) (these patterns agree well with the numerically calculated ones shown in the Extended Data Fig.~2). \rev{The modulation of the refractive index in these experimentally created quasicrystal lattices, under the peak intensity of the writing beam which is fixed to be $40~ \text{W/m}^2$ and the applied electric field $E=300~\text{V/mm}$,
is $\delta n=4.4~\times 10^{-4}$.}
\begin{figure}[htb]
\centering
\includegraphics[width=\columnwidth]{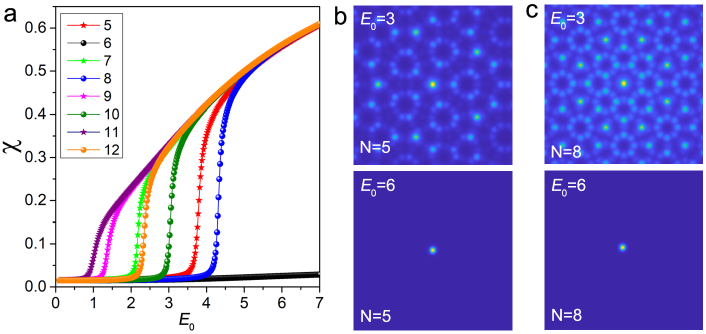}
\caption{ \setlength{\baselineskip}{10pt} {\bf Form factors and profiles of linear eigenmodes supported by quasicrystals.}  ({\bf a}) The dependence of the form factor (inverse width) of the eigenmode with largest $\beta$ on electric field $E_0$, for quasicrystals with different orders of discrete rotational symmetry $N$. For the values of d.c. field $E_0$ well above the localization threshold all curves $\chi(E_0)$ approach the numerically found asymptotic curve $\chi_{\rm as}\approx-0.2138+0.3162\sqrt{E_0}$. The curves with stars and spheres correspond to quasicrystals with odd and even discrete rotational symmetries, respectively. ({\bf b}) and ({\bf c}) show the examples of the mode profiles $|w({\bm r})|^2$ with the largest $\beta$ for $E_0<E_0^{\rm LDT}$ (upper panel) and $E_0>E_0^{\rm LDT}$ (lower panel) for quasicrystal with ({\bf b}) $N=5$ and ({\bf c}) $N=8$. The window shown in ({\bf b}) and ({\bf c}) is $-20 \le x \le 20$ and $-20 \le y \le 20$.
}
\label{fig:two}
\end{figure} 

The field of a stationary probe beam (i.e., of an eigenmode) propagating in the quasicrystalline structure covering the whole transverse face of the sample is described by a solution of the Eq.~(\ref{equation1}) of the form $\psi({\bm r},z)=w({\bm r})e^{i\beta z}$, where $w({\bm r})$ is the transverse profile of the mode, solving the eigenvalue problem $Hw=-\beta w$, and $\beta$ is the propagation constant. \rev{The eigenvalue problem was solved using the finite-difference method.} To characterize the degree of the localization of such eigenmode one can use the integral form factor \rev{(alias inverse participation ratio)} $ \chi=(\iint |\psi|^{4}d^2{\bm r})^{1/2}/U$, where $ U=\iint|\psi|^{2}d^2{\bm r}$ and the integration is over the transverse area of the quasicrystal, is the energy flow. The form factor is inversely proportional to the average width of the mode: the larger the value of $\chi$, the stronger the localization. The dependence of the form factor of the most localized mode supported by the photonic quasicrystal, i.e., of the mode with the largest $\beta$, on $E_0$ for different values of $N$ is shown in Fig.~\ref{fig:two}(a). The central result of this work is that we observe that for quasicrystals of any rotational symmetry, prohibited by the crystallographic restriction theorem, there exists a critical depth of the potential, defined by the applied d.c. field and denoted below as $E_N^{\rm{LDT}}$ ("LDT" stands here for the localization-delocalization transition), above which, i.e., at $E_0>E_N^{\rm{LDT}}$, at least one of the guided modes becomes spatially localized, while at $ E_0<E_N^{\rm{LDT}}$ all eigenmodes of the Hamiltonian $H$ are extended. Since the transition between strongly localized and delocalized states occurs within the interval of $E_0$ values that have finite widths, the critical value $E_N^{\rm{LDT}}$ is defined as the point, where the respective dependence $\chi (E_0)$ changes its slope.

Within each group of odd and even order symmetries the critical depths $E_N^{\rm{LDT}}$ decrease rapidly with the increase of the symmetry order, i.e., $E_5^{\rm{LDT}}>E_7^{\rm{LDT}}>\cdots$ and $E_8^{\rm{LDT}}>E_{10}^{\rm{LDT}}>\cdots$ [see curves marked respectively by stars and spheres in Fig.~\ref{fig:two}(a)]. \rev{This separation into two sequences of symmetries corroborates with different properties of dihedral groups of odd and even orders \cite{Cameron1998}. In particular, quasicrystals obey (lack) inversion symmetry along each of the symmetry axes for even (odd) $N$, which is expected to affect the light propagation. We have also found that the positions of the LDT thresholds on the energy scale correlate with the effective inhomogeneity of the refractive index, which can be characterized by the deviation of the integral refractive index at the center of the quasicrystal from its asymptotic value on the periphery (this property can be quantified by the filling fraction $f_N$; it is investigated in Supplementary Information).} For instance, we have found that the LDT threshold is a monotonically decreasing function of the filling fraction $f_N$, as shown in Fig.~S3(c) of the Supplementary Information. Meantime, for quasicrystals belonging to dihedral groups of even and odd orders the critical depths are not strictly alternating, for example, $E_8^{\rm{LDT}}>E_5^{\rm{LDT}}>E_{10}^{\rm{LDT}}>E_{12}^{\rm{LDT}}>\cdots$. Notice that for the structure with $N=6$ that is consistent with the crystallographic restriction theorem, i.e., is exactly periodic, no LDT is observed: all eigenmodes remain delocalized for any $E_0$ value that is manifested in low values of the form factor $\chi$. 

\rev{To gain a better theoretical understanding of the LDT phenomenon, we performed calculations of the density of states (DOS) in photonic quasicrystals with different symmetries. These calculations were conducted both above and below the LDT point, as shown in Figure S1 in Supplementary Information. Above the LDT point, specifically for applied field $E_0 > E_N^{\text{LDT}}$, the distribution of the DOS (DOS($\beta$)) in the quasicrystals of all orders exhibits a discrete-like nature with multiple spikes and extended regions of the spectrum where the DOS is nearly zero. In this case, the eigenmode with the largest value of eigenvalue $\beta$ is well separated from the other modes. In contrast, below the LDT point, with applied field  $E_0 < E_N^{\text{LDT}}$, the DOS($\beta$) distribution appears nearly continuous. In this regime, there are numerous delocalized modes that are distributed in a nearly uniform manner across the spectrum.}

 Yet another interesting observation is the existence of \rev{two limits in the dependencies $\chi(E_0)$ shown in Fig.~\ref{fig:two} (a). The first one is} a "saturation limit" for the form factors of the fundamental eigenmodes guided by quasicrystals of different symmetries: well above the critical potential depth the form factors of modes in structures with different $N$ asymptotically approach the curve, which is well approximated by the formula $\chi_{\rm as}\approx-0.2138+0.3162\sqrt{E_0}$, indicating on the fact that characteristic localization radii of the well-localized fundamental modes are practically independent of the symmetry order $N$.  \rev{This independence of the form factor on the symmetry order can be explained by the isotropic properties of the central maximum of the refractive index (Supplementary Information). The second limit is the minimal LDT threshold achieved at $E_0\approx 0.4572$ when $N\to\infty$. The existence of this minimum value of LDT can be understood by observing  that  $ \lim_{N\to\infty}I_N(r,\varphi)=[AJ_0(2r)]^2$  where $J_0(\cdot)$ is the zero-order Bessel function of the first kind (Supplementary Information). In this limit, the propagation problem is reduced to the existence (or non-existence) of bound states in the respective confining optical potential $-E_0[AJ_0(2r)]^2/(1+[AJ_0(2r)]^2)$ which tends to zero at $r\to\infty$. However, this decay is very slow and the known results~\cite{LL, Simon} on the existence of at least one bound state at any $E_0$, are not applicable to it. Instead, in Supplementary Information we show that for $E_0$ small enough such potential cannot sustain guided modes, i.e., guidance can be enabled only by the bias fields of finite amplitudes. }  

We emphasize that localized eigenmodes reported here are not defect modes, which one may expect to find in deep potential minima - in contrast, these states are enabled by interference. In order to illustrate this, in Fig.~\ref{fig:one}(b) we show representative profiles of the localized fundamental modes and their "energies" $-\beta$ (green dashed lines) with respect to the optical potential. One observes that for both $N=5$ and $N=8$ structures, the $-\beta$ energy level crosses multiple local potential minima (the behavior that clearly contrasts with that of the defect modes), but the fundamental modes remain strongly localized in the vicinity of the central minimum and do not undergo diffraction due to tunneling to the nearest local minima.

In Fig.~\ref{fig:two}(b) and (c) we compare the profiles of the fundamental modes in the quasicrystals with $N=5$ having $E_5^{\rm{LDT}}\approx 3.6$ and $N=8$ having $E_8^{\rm{LDT}}\approx 4.1$, for potential depths below and above respective $E_N^{\rm{LDT}}$. The modes for $E_0=3<E_{5,8}^{\rm{LDT}}$ are delocalized (only central regions are shown in the figure, while the actual calculation window is much larger), while the modes at $E_0=6>E_{5,8}^{\rm{LDT}}$ are localized practically on one central spot of the potential. This is the case for all quasicrystal lattices with $E_0$ value well above the critical depth, since form factors of fundamental modes in this limit approach practically the same value $\chi_{\rm as}(E_0)$.

\begin{figure}[htb]
\centering
\includegraphics[width=\textwidth]{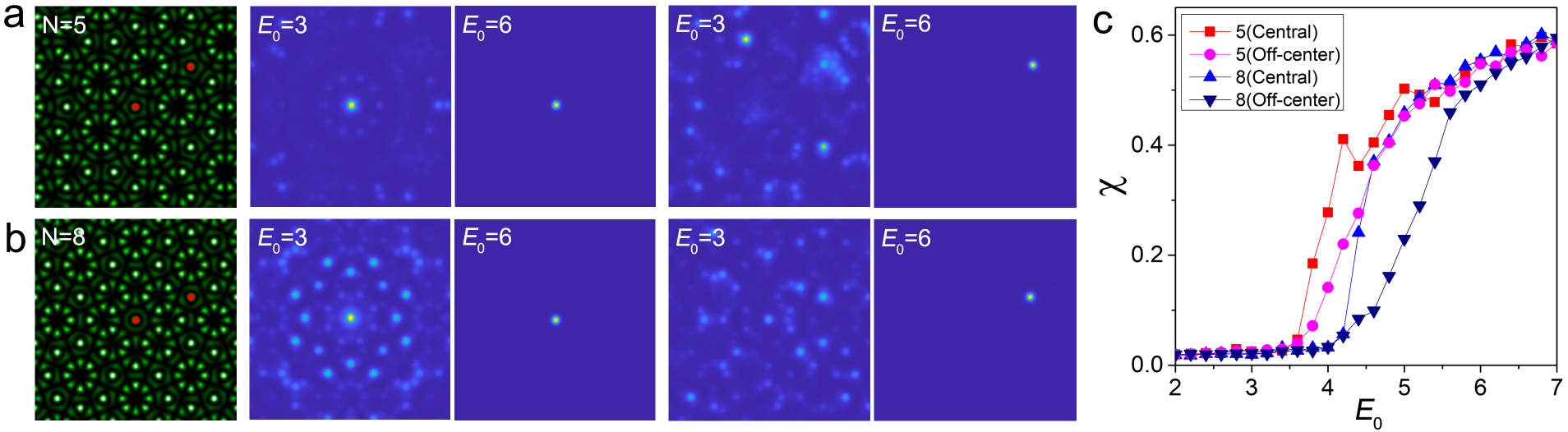}
\caption{ \setlength{\baselineskip}{10pt} {\bf Propagation of input Gaussian beams and their form factors.} Simulated propagation of an input Gaussian beam in five-fold \textbf{(a)} and eight-fold \textbf{(b)} symmetric quasicrystals. The red dots in left lattice images in ({\bf a}) $(N=5)$ and ({\bf b}) $(N=8)$ indicate the location of the central and off-center Gaussian excitations. Corresponding output intensity distributions after propagation distance $z=50$ cm in the quasicrystal are shown in the second and third columns for central excitation and in the fourth and fifth columns for off-center excitation for the applied electric field $E_0=3$ (second and fourth column) and $E_0=6$ (third and fifth column). The window shown in ({\bf a}) and ({\bf b}) is $-20 \le x,y \le 20$. ({\bf c}) Form factor of the output light field extracted from the propagation simulation at a much larger distance $z=50~\text{cm}$ versus $E_0$ for both central and off-center input beams.
}
\label{fig:three}
\end{figure}

Simulations of direct propagation within the framework of Eq.~(\ref{equation1}), \rev{using split-step FFT method},  with an input Gaussian beam $\psi({\bm r},0)=\exp(-|{\bm r}|^2/r_0^2)$ of the width $r_0=1$ that covers roughly one local maximum of the induced quasicrystal lattice, are reported in second and third panels of Fig.~\ref{fig:three}(a) and (b). As predicted by the analysis of the eigenvalue problem $Hw=-\beta w$, for $E_0<E_N^{{\rm LDT}}$, the beam quickly diffracts 
(second column of Fig.~\ref{fig:three}), while for $E_0$ above $E_N^{{\rm LDT}}$, the beam remains well-localized at all distances (third column of Fig.~\ref{fig:three}). Importantly, this observation does not depend on the position of the initial Gaussian excitation at the input face of the crystal. Namely, for the off-center incidence described by $\psi({\bm r}-\bm{r}_{in})$ $({\bm r}_{in}=11.3 {\bm i}+8.1 {\bm j}$ \rm for (a), and $ {\bm r}_{in}=11.1 {\bm i}+4.6 {\bm j}$ for (b); they were chosen to be one of the $N$ local maxima on a ring of a certain radius in the lattice; see also Fig.~S4 in Supplementary Information for the illustration), the beam also shows diffraction or localization depending on whether the $E_0$ value is below or above the critical value $E_N^{{\rm LDT}}$ (see fourth and fifth column in Fig.~\ref{fig:three}), just as for the central excitation. Indeed, the integral form factor $\chi$, calculated at the distance $z=500$ (corresponding to a physical distance $z= 50$ cm), illustrates that localization-delocalization transition value $E_N^{{\rm LDT}}$ is practically the same for the central and off-center excitations [Fig.~\ref{fig:three}(c); for more results on off-center excitation see Fig.~S2 and Fig.~S4 of Supplemental Information].
 
\begin{figure}[htb]
\centering
\includegraphics[width=\textwidth]{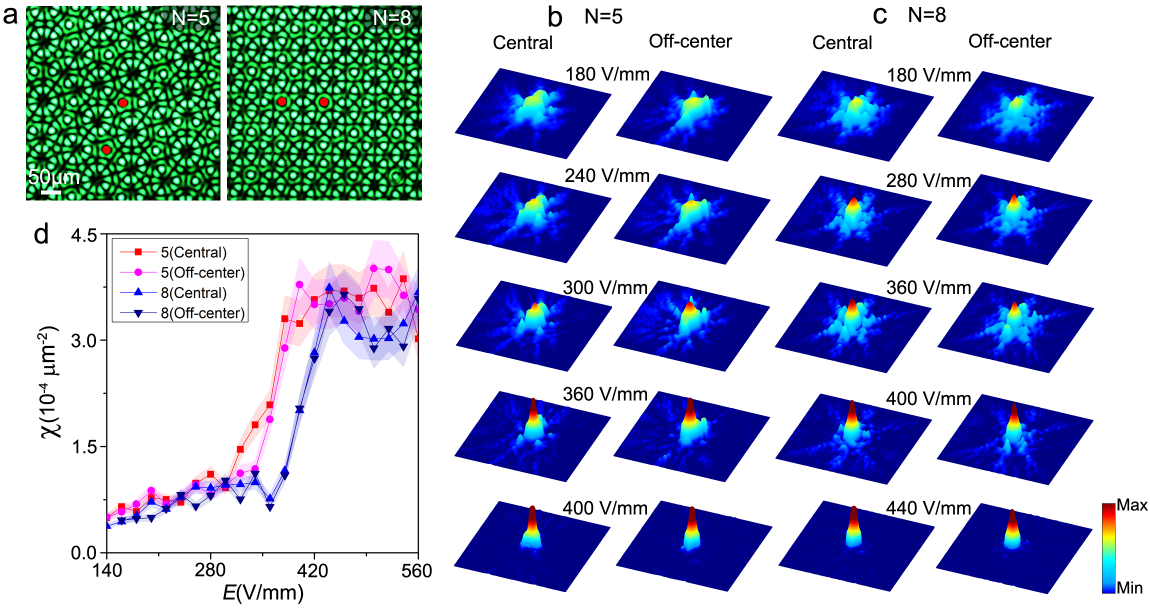}
\caption{ \setlength{\baselineskip}{10pt} {\bf Experimental observation of light localization in quasicrystals.} ({\bf a}) Experimentally realized quasicrystal lattices with $N=5$ and $N=8$. The red dots indicate the locations where the probe beam is launched. ({\bf b}) Observed output intensity distributions of probe beam after propagating in quasicrystal lattices with $N=5$, illustrating localization-delocalization transition with the increase of the applied electric field for both central (left column) and off-center excitations (right column). ({\bf c}) The same as ({\bf b}) but for quasicrystals with $N=8$. The applied field $E$ measured in V/mm, is indicated between the respective output profiles. \rev{In ({\bf b}) and ({\bf c}), the distributions are shown within the window of
$400~\mu\text{m} \times 400~\mu \text{m}$.} ({\bf d}) Experimentally measured form factor at the output facet of the quasicrystal versus applied field for central and off-center excitations. The shaded area represents the uncertainties of the form factor introduced during the lattice creation and signal beam measurement.
}
\label{fig:four}
\end{figure}

For the experimental observation of light localization we first employ representative members of photonic quasicrystals family with odd $N=5$ and even $N=8$ discrete rotational symmetries, depicted in Fig.~\ref{fig:four}(a). To probe light propagation in them, an extraordinarily polarized signal beam was focused on the input facet of the quasicrystal lattice. The signal beam had about $22~\mu \textrm{m}$ in diameter and it was coupled into a selected local refractive index maximum of the quasicrystal. The selected local maximum can be located in the rotational center of the quasicrystal, or it can be launched at the off-center position, as individually indicated with the red dots in Fig.~4(a). To ensure that the input beam does not distort the induced refractive index profile and that it propagates in the crystal in the linear regime, \rev{the bias field was turned off after the lattice was prepared, and the power of signal beam was taken approximately $10~ \text{nW}$, nearly $10^3$ times lower than the power of the lattice-creating beam. }



Experimental evidence of the light localization in quasicrystals for both central and off-center excitation conditions is presented in Fig.~\ref{fig:four}, where we compare output patterns for the probe beam after propagation through the five-fold ($N=5$) and eight-fold ($N=8$) symmetric $2~\textrm{cm}$-long quasicrystals for different applied electric fields $E$. \rev{More results on quasicrystal lattices with other $N$
are given in Extended Data Fig.~\ref{fig:LDT2}.}
As shown in Figs.~\ref{fig:four}(b) and \ref{fig:four}(c), for a fixed $N$, a relatively sharp LDT occurs when $E$ exceeds certain critical value $E_N^{\rm LDT}$, indicating a clear transition from the beam diffraction to spatial localization. In the experiment, we measured  $E_5^{\rm LDT} \approx 300~\text{V/mm}$ and $E_8^{\rm LDT} \approx 360~\text{V/mm}$ for quasicrystals featuring five- and eight-fold rotational symmetries, respectively. Thus, as compared in Figs. \ref{fig:four}(b) and \ref{fig:four}(c), when $E<E_N^{\rm LDT}$, the light in the quasicrystal lattices notably diffracts upon propagation and expands across multiple local maxima in the vicinity of the excitation point. When $E>E_N^{\rm LDT}$, diffraction is clearly arrested and a localized spot is observed at the output. As predicted by the numerical analysis and confirmed here experimentally, the critical applied field $E_N^{\rm LDT}$ is nearly the same for the central and off-center excitation conditions [Fig.~\ref{fig:four}(d)].

\begin{figure}[htb]
\centering
\includegraphics[width=\columnwidth]{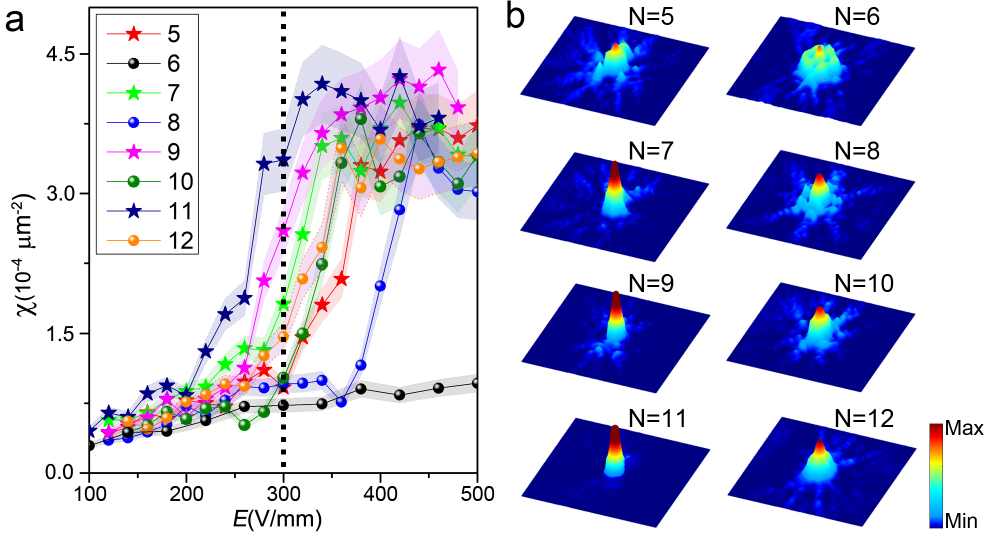}
\caption{ \setlength{\baselineskip}{10pt} {\bf Experimental results for different orders of the point rotational symmetry of the quasicrystals.} Experimentally measured form factor versus applied field ({\bf a}), and the observed output intensity distributions for the probe beam at $E=300$~V/mm (indicated in (a) by the vertical dashed line) ({\bf b}), for quasicrystal lattices with $N$ increasing from 5 to 12. In all the cases shown here, central excitation is used. The shaded area in ({\bf a}) represents the uncertainties of the form factor,\rev{ and in ({\bf b})  the distributions are shown within the window of $400~\mu\text{m} \times 400~\mu \text{m}$.} 
}
\label{fig:five}
\end{figure}  

\rev{It is important to note that the experimentally observed localizations in quasicrystal lattices are due to the interference effect, i.e., it is determined by global, rather than local, symmetry properties of the underlying structures. To confirm that no additional disorder or defects were introduced into our optically induced quasicrystal lattices, which could lead to light localization, we conducted a comparison experiment in the commensurate case ($N=6$) for which the optically induced structure is exactly periodic. In this case, we observed significant diffraction of light for any position of the probe beam (Extended Data Fig.~3). Since we used the same setup and method to induce all quasicrystal lattices for varying $N$ (they differ only in the employed phase masks), this ensures that disorder and defects are absent in the lattices for any other $N$ values. Besides, to ensure the reproducibility of the results and to eliminate the possible influence of any anisotropy in the response of photorefractive crystals, we performed the experiments five times for each $N$ value. Before each experiment, the previously created structures were erased, and the quasicrystal lattice was rewritten to a different location across the sample (this was achieved by displacement of the mask; see \textbf{Methods}). We then repeated measurements of the output patterns. Therefore, each point in Figs.~\ref{fig:four} (d) and \ref{fig:five} (a) represents the average of five measurements, and this guarantees that any possible disorder of defects are ruled out as factors affecting localization in specific realizations of the lattice.}

Finally, we experimentally study the impact of the symmetry order $N$ of the quasicrystal on the light localization by comparing the respective form factors $\chi$ for the output probe beams versus $E$, for $N$ increasing from $N=5$ to $12$. As shown in Fig.~\ref{fig:five}(a), for all orders of discrete rotational symmetry of the crystal except for $N=6$ (corresponding to the periodic lattice), the dependence of $\chi$ on $E$ exhibits a clear jump from lower to higher value after undergoing a rather narrow transition regime, and the transition point clearly decreases with the increase of the order $N$ (if one considers odd and even $N$ values separately), which well agrees with the numerical results presented in Fig.~\ref{fig:two}. \rev{As one can see, the experimental measurements reveal saturation of the form factor around $E=400$ V/mm, where we estimate induced refractive index as $\delta n=5.87 \times 10^{-4}$.}
In Fig.~5(b), the output intensity distributions are presented for quasicrystals with different rotational point symmetries at the fixed electric field $E=300$~{V/mm}, which is substantially lower than the critical field $E_5^{\rm LDT}$ for five-fold symmetric quasicrystal. Accordingly, for $N=5$ the probe beam experiences notable diffraction upon propagation. Diffraction is also observed for $N=8$, $10$ and $12$. For quasicrystal with $N=7$ the electric field is near the critical value at which transition to localization occurs, so the tendency to suppression of diffraction is obvious. Finally, for crystals with $N=9,11$, the selected electric field is larger than the critical value, and one observes well-pronounced localization on the entire length of the sample. Note that, in our experiment, the use of the translation stage with a mounted CCD-camera (Extended Data Fig.~1) allows recording of the light intensity at every distance $z$ inside the sample.
Meantime, our  experimental results do not reproduce the law $\chi_{\rm as}(E_0)$, shown in Fig.~\ref{fig:two} (a). Instead, the increase of $E_0$ results in the saturation of $\chi$ to a constant. We attribute this disagreement of the numerical and experimental asymptotic behaviors to the fact that the crystal cannot provide indefinitely high refractive index contrast: nonlinear effects start playing the role at sufficiently strong values of the field $E_0$.

\rev{In conclusion, we reported the first investigation and observation of the impact of discrete rotational symmetry on the localization-delocalization transition in a photonic quasicrystal. By continuously tuning parameters, we found that the critical depth of quasicrystals  
controlling the localization-delocalization transition decreases with the increase of the order of the discrete rotational symmetry.} The observation reported in this work clarifies fundamental aspects of the evolution of linear excitations in wave systems with quasicrystal structure and may shed light on the explanation of localization phenomena in aperiodic photonic crystals and photonic crystal fibers, phononic systems, and Bose–Einstein condensates held in optically induced quasicrystal lattices~\cite{Sanchez2005,Viebahn2019,Gautier2021}. The phenomenon of light localization in quasicrystals may be employed for the design of micro-lasers without the need for conventional laser cavity~\cite{moirelaser}, may be used to enhance the nonlinear parametric interactions of light waves~\cite{shg}, and in cavity quantum electrodynamics.

\section*{METHODS}
 \paragraph*{\bf Experimental setup.} The experimental setup is sketched in Extended Data Fig.~1. \rev{We used the technique of computer-generated holography (CGH) to produce the desired quasicrystal lattice. Initially, the phase information of the targeted lattice, which corresponds to the interference wave field of $N$ pairs of counterpropagating plane waves, was encoded into a phase-only Spatial Light Modulator(SLM) that had a resolution of 1920×1200 pixels controlled by a computer. An example phase diagram  ($N=5$) encoded into the SLM is presented in the inset of Extended Data Fig. 1, and more diagrams are presented in the Supplementary Information. The interference wave field was then reconstructed by illuminating the SLM with a cw-laser with $\lambda=532~\text{nm}$ and ordinary polarization. It should be noted that, although the reconstructed wave field appears visually similar to the target quasicrystal lattices, it cannot be directly “written” into the SBN crystal as it does not form a well-nondiffracting light beam. To overcome this limitation, we converted the wave field into the wavevector domain, filtered out any unwanted components (using a Fourier Mask placed at the Fourier plane of the 4f optical system, which allowed only the first-order diffraction pattern to pass),   and then transformed it back to the space domain. This process ensures the creation of a smooth and well-nondiffracting wave field throughout the 2 cm long of sample. }
 
\rev{A signal light with wavelength 633~nm and extraordinary polarization was used to probe the propagation dynamics of light in photonic quasicrystal lattices. A translation stage equipped with a CCD camera was used to record the intensity distribution of the probe beam at different locations inside the sample.}

\rev{In our experiments involving off-center excitations, we selected one of the $N$ local maxima on a ring with a certain radius in the lattice as the excitation point.  We usually chose any of the $N$ local lattice maxima on the first, second, or third rings with radii ranging from tens to a few hundreds of $\mu \text{m}$. As the input probe beam has a diameter of approximately $22~\mu \text{m}$, while the sample has a transverse dimension $5\text{mm} \times 5\text{mm}$, we made sure that the selected excitation points were sufficiently far away from the boundaries of the finite-size structure. As a result, the impact of the boundary effects can be safely ignored.
}
 \setcounter{figure}{0}
 \renewcommand\figurename{Extended Data Fig.}
 
\begin{figure}[ht]
 \centering
 \includegraphics[width=0.9\textwidth]{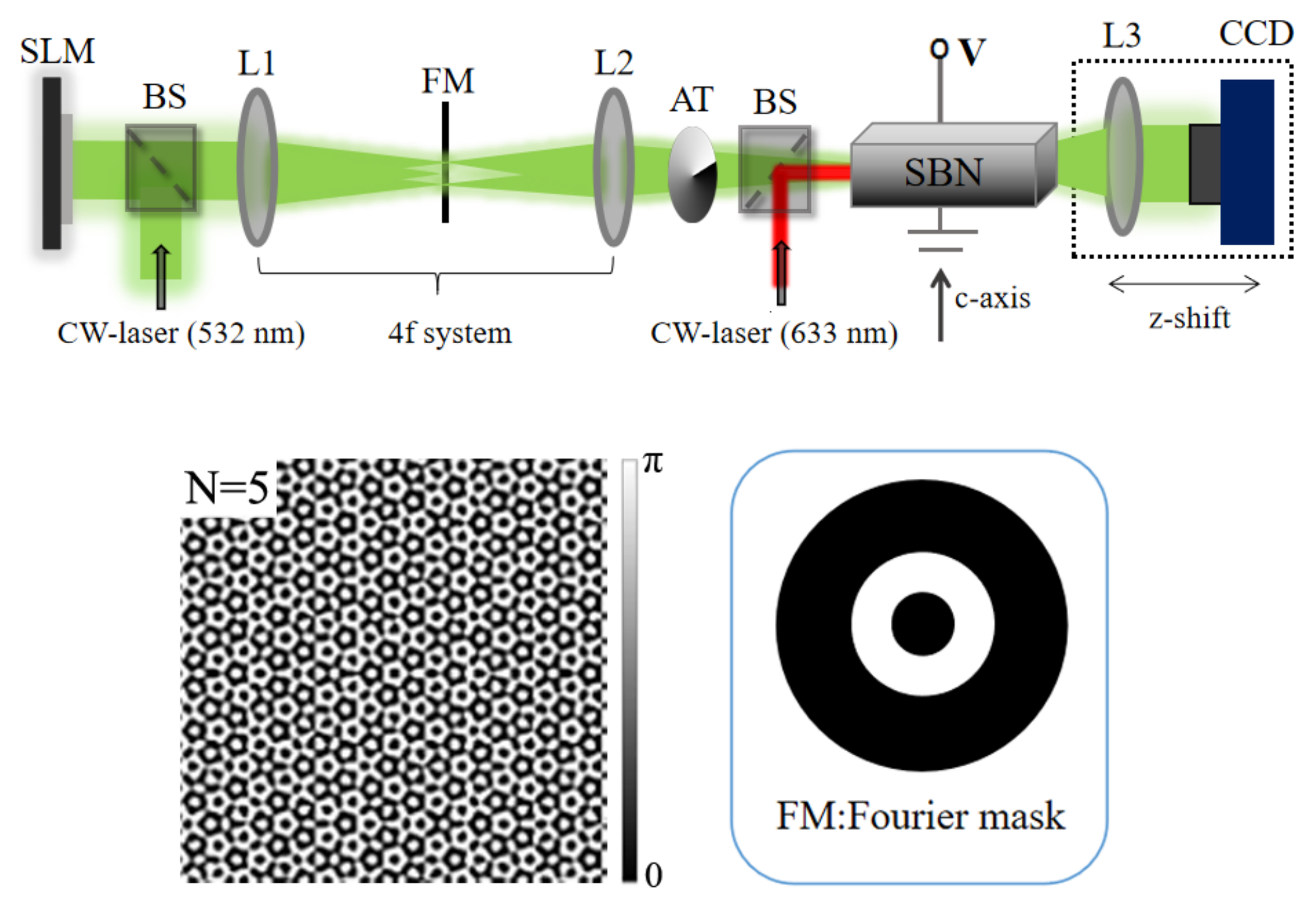}
\caption{Experimental setup. SLM, spatial light modulator; BS, beam splitter; L, lens; {FM, Fourier mask; AT,variable attenuator}; SBN, strontium barium niobate crystal; CCD, charged-coupled device. Bottom-left, the phase diagram for N = 5; bottom-right, the structure of Fourier mask.
}
 \label{fig:experiment}
 \end{figure}

 \begin{figure}[ht]
 \centering
 \includegraphics[width=0.9\textwidth]{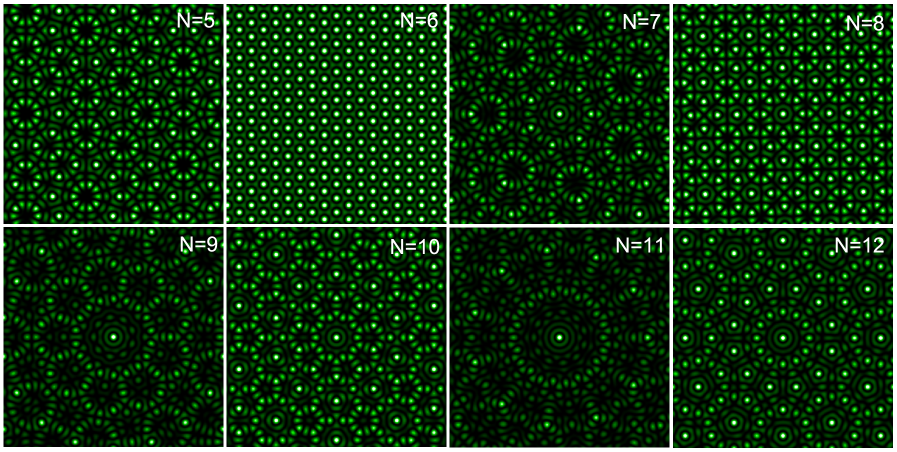}
\caption{Numerically calculated $N$-fold symmetric quasicrystal patterns with $N = 5,~7,~8..,12$, as well as the periodic pattern with $N=6$, for $k=2$ and $A^2=2.24$.
}
 \label{fig:simulation patterns}
 \end{figure}

 \begin{figure}[ht]
\centering
\includegraphics[width=\columnwidth]{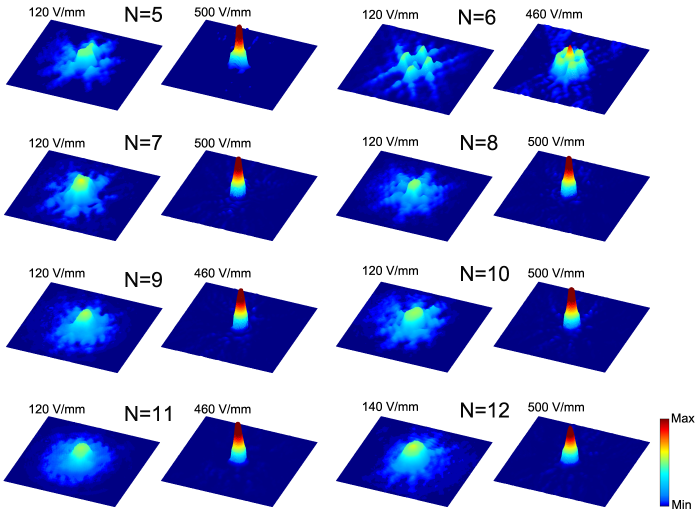}
\caption{Experimentally observed delocalized output intensity distributions for probe beam observed below LDT point (at low values of the applied electric field) and localized distributions observed above LDT point (at sufficiently high values of the electric field), for $N=5, 7-12$. At $N=6$ the field is delocalized at all amplitudes of the electric field. The distributions are shown within the window of 400 $\mu$m $\times$  400 $\mu$m.
}
\label{fig:LDT2}
\end{figure} 

\paragraph*{\bf Data Availability}
The data that support the findings of this study are available from the corresponding author F. Y. upon reasonable request.
 
\paragraph*{\bf Acknowledgments}
  P. W., Q. F. and F. Y. acknowledge the support of the Natural Science Foundation of Shanghai (No.19ZR1424400), and Shanghai Outstanding Academic Leaders Plan (No. 20XD1402000).  
 V. V. K. acknowledges financial support from the Portuguese Foundation for Science and Technology (FCT) under Contracts PTDC/FIS-OUT/3882/2020 and UIDB/00618/2020.
 Y. V. K. acknowledges funding by the research Project No. FFUU-2021-0003 of the Institute of Spectroscopy of the Russian Academy of Sciences. We would like to express our gratitude to the anonymous referees for their valuable suggestions regarding the filling fractions and structure factors used to elucidate the order of the quasicrystals in explaining the threshold LDT.

\paragraph*{\bf Author contributions} 
P. W and Q. F  contribute equally to this work. All authors contribute significantly to the work.

\paragraph*{\bf Competing interests} 
The authors declare no competing interests.

\end{document}